\documentclass[%reprint,
superscriptaddress,
nofootinbib,
 amsmath,
 amssymb,
 aps,
prd,
floatfix,
showkeys
]{revtex4-2}
\usepackage{longtable}
\usepackage{aas_macros}%for bib
\usepackage{graphicx}% Include figure files
\usepackage{dcolumn}% Align table columns on decimal point
\usepackage{siunitx}
\usepackage{bm}% bold math
\DeclareSIUnit \year {yr}
\renewcommand*{\i}{\mathrm{i}}

\begin{document}
%\graphicspath{{./}}

\preprint{APS/123-QED}

\title{Detection of Partial Coherence due to Multipath Propagation for FRB 20220413B with CHIME/FRB}% Force line breaks with \\
%\thanks{A footnote to the article title}%

\author{Zarif Kader}
  \affiliation{Department of Physics, McGill University, 3600 rue University, Montr\'eal, QC H3A 2T8, Canada}
  \affiliation{Trottier Space Institute, McGill University, 3550 rue University, Montr\'eal, QC H3A 2A7, Canada}
\author{Evan Davies-Velie}
  \affiliation{Department of Physics, McGill University, 3600 rue University, Montr\'eal, QC H3A 2T8, Canada}
  \affiliation{Trottier Space Institute, McGill University, 3550 rue University, Montr\'eal, QC H3A 2A7, Canada}
\author{Matt Dobbs}
  \affiliation{Department of Physics, McGill University, 3600 rue University, Montr\'eal, QC H3A 2T8, Canada}
  \affiliation{Trottier Space Institute, McGill University, 3550 rue University, Montr\'eal, QC H3A 2A7, Canada}  
\author{Afrokk Khan}
  \affiliation{Department of Physics, McGill University, 3600 rue University, Montr\'eal, QC H3A 2T8, Canada}
  \affiliation{Trottier Space Institute, McGill University, 3550 rue University, Montr\'eal, QC H3A 2A7, Canada}
\author{Calvin Leung}
  \affiliation{Miller Institute for Basic Research, Stanley Hall, Room 206B, Berkeley, CA 94720, USA}
  \affiliation{Department of Astronomy, University of California, Berkeley, CA 94720, United States}
\author{Robert Main}
  \affiliation{Department of Physics, McGill University, 3600 rue University, Montr\'eal, QC H3A 2T8, Canada}
  \affiliation{Trottier Space Institute, McGill University, 3550 rue University, Montr\'eal, QC H3A 2A7, Canada}  
\author{Kiyoshi W. Masui}
  \affiliation{MIT Kavli Institute for Astrophysics and Space Research, Massachusetts Institute of Technology, 77 Massachusetts Ave, Cambridge, MA 02139, USA}
  \affiliation{Department of Physics, Massachusetts Institute of Technology, 77 Massachusetts Ave, Cambridge, MA 02139, USA}
\author{Kenzie Nimmo}
  \affiliation{Center for Interdisciplinary Exploration and Research in Astronomy, Northwestern University, 1800 Sherman Avenue, Evanston, IL 60201, USA }
  \affiliation{MIT Kavli Institute for Astrophysics and Space Research, Massachusetts Institute of Technology, 77 Massachusetts Ave, Cambridge, MA 02139, USA}
\author{Ue-Li Pen}
  \affiliation{Institute of Astronomy and Astrophysics, Academia Sinica, Astronomy-Mathematics Building, No. 1, Sec. 4, Roosevelt Road, Taipei 10617, Taiwan}
  \affiliation{Canadian Institute for Theoretical Astrophysics, 60 St.~George Street, Toronto, ON M5S 3H8, Canada}
  \affiliation{Canadian Institute for Advanced Research, 180 Dundas St West, Toronto, ON M5G 1Z8, Canada}
  \affiliation{David A.\ Dunlap Department of Astronomy and Astrophysics, 50 St. George Street, University of Toronto, ON M5S 3H4, Canada}  
  \affiliation{Perimeter Institute of Theoretical Physics, 31 Caroline Street North, Waterloo, ON N2L 2Y5, Canada}
\author{Mawson Sammons}
  \affiliation{Department of Physics, McGill University, 3600 rue University, Montr\'eal, QC H3A 2T8, Canada}
  \affiliation{Trottier Space Institute, McGill University, 3550 rue University, Montr\'eal, QC H3A 2A7, Canada}

% Unique acks:
\newcommand{\allacks}{ We acknowledge that CHIME is located on the traditional, ancestral, and unceded territory of the Syilx/Okanagan people. We are grateful to the staff of the Dominion Radio Astrophysical Observatory, which is operated by the National Research Council of Canada. CHIME operations are funded by a grant from the NSERC Alliance Program and by support from McGill University, University of British Columbia, and University of Toronto. CHIME was funded by a grant from the Canada Foundation for Innovation (CFI) 2012 Leading Edge Fund (Project 31170) and by contributions from the provinces of British Columbia, Québec and Ontario. The CHIME/FRB Project was funded by a grant from the CFI 2015 Innovation Fund (Project 33213) and by contributions from the provinces of British Columbia and Québec, and by the Dunlap Institute for Astronomy and Astrophysics at the University of Toronto. Additional support was provided by the Canadian Institute for Advanced Research (CIFAR), the Trottier Space Institute at McGill University, and the University of British Columbia. The CHIME/FRB baseband recording system is funded in part by a CFI John R. Evans Leaders Fund award to IHS. M.D. is supported by a CRC Chair, NSERC Discovery Grant, CIFAR, and by the FRQNT Centre de Recherche en Astrophysique du Qu\'ebec (CRAQ). K.W.M. is supported by NSF Grant Nos. 2008031, 2510771 and holds the Adam J. Burgasser Chair in Astrophysics. K.N. acknowledges support by NASA through the NASA Hubble Fellowship grant \# HST-HF2-51582.001-A awarded by the Space Telescope Science Institute, which is operated by the Association of Universities for Research in Astronomy, Incorporated, under NASA contract NAS5-26555. U.P. is supported by the Natural Sciences and Engineering Research Council of Canada (NSERC), Canadian Institute for Advanced Research (CIFAR), Ontario Research Fund Research Excellence and by the AMD AI Quantum Astro. }
\date{December 12, 2025}% It is always \today, today,
             %  but any date may be explicitly specified

\begin{abstract}
Fast radio bursts (FRBs) are a $\sim$ millisecond-long transient phenomenon that propagate across extragalactic distances and are effectively a point source. Radio wave propagation through inhomogeneous distributions of plasma can act as a lens, generating multiple images of the emitted electric field. A lens can produce images of a point source where the phase of the electric field between images remains coherent when observed by a radio telescope. FRB 20220413B shows a complicated pulse structure with time separated components that may be image copies of the main components due to plasma lensing. We perform several analyses to determine if FRB 20220413B is consistent with expectations of a plasma lensed FRB. We analyze and fit the morphology of the burst to a plasma lens model and find consistency in the spectro-temporal profile but not the observed flux. Using the complex-valued channelized voltage data from the CHIME telescope, we perform a time-lag correlation analysis and report correlation signatures present in the electric field of FRB 20220413B. We find that there exists an excess correlation signature only in absolute power and not in phase. We perform a frequency-lag correlation analysis on the spectra of all subcomponents of the burst and find a consistent scintillation bandwidth across all components. We find the scintillation bandwidth is consistent with expectations of scattering due to the Milky Way. We interpret this as all burst components propagating through the same scintillation screen located in the Milky Way, which would generate the excess variance signature observed, even in the absence of phase coherence between burst components. We find that while the burst morphology can be modeled by a plasma lens, the coherent signature present in the time-lag correlation is consistent with the expectations of a common scattering screen, but not coherent plasma lensing.
\end{abstract}
%% Keywords should appear after the \end{abstract} command. 
%% The AAS Journals now uses Unified Astronomy Thesaurus concepts:
%% https://astrothesaurus.org
%% You will be asked to selected these concepts during the submission process
%% but this old "keyword" functionality is maintained in case authors want
%% to include these concepts in their preprints.

%% From the front matter, we move on to the body of the paper.
%% Sections are demarcated by \section and \subsection, respectively.
%% Observe the use of the LaTeX \label
%% command after the \subsection to give a symbolic KEY to the
%% subsection for cross-referencing in a \ref command.
%% You can use LaTeX's \ref and \label commands to keep track of
%% cross-references to sections, equations, tables, and figures.
%% That way, if you change the order of any elements, LaTeX will
%% automatically renumber them.
%%
%% We recommend that authors also use the natbib \citep
%% and \citet commands to identify citations.  The citations are
%% tied to the reference list via symbolic KEYs. The KEY corresponds
%% to the KEY in the \bibitem in the reference list below. 

\maketitle

%\maketitle
%\newpage

\section{Introduction} \label{sec:intro}
The radio waves emitted from distant astrophysical processes can propagate through media in space that can act like a lens. These astrophysical lenses can vary in origin from stars \cite{Connor2023}, black holes \cite{Munoz2016,Eichler2017,Katz2020}, or discrete distributions of plasma \cite{Fiedler1987,Clegg1998}. Observation of these lenses provides insight into their physical properties. For example, gravitational lenses provide insight into the nature of dark matter, as it may only affect light through the gravitational field. Plasma lenses are generated by inhomogeneities in the plasma electron density and directly constrain the environments around active galactic nuclei and compact objects. These astrophysical lenses also provide information about the source that emitted the radio waves. A point source emitter has a unique property where the phase of the original field can be preserved after propagating through a lens. The lens can create image copies of the original field with a distinct phase relation, which can be measured and detected by a radio telescope \cite{Kader2024}. Fast radio bursts (FRBs) are radio transient events that are approximately a millisecond long in duration \cite{Lorimer2007} originating from extragalactic distances \cite{Bochenek2020}. To date, the emission mechanism of FRBs is not known \cite{Petroff2022}. It is therefore important to understand the effects of propagation on the signals we observe. In doing so, we can identify and distinguish the properties of the unknown emission mechanism from propagation effects such as astrophysical lensing.

While the emission process is not fully understood, an observed link to magnetars \cite{Bochenek2020} suggests compact objects may be the originator of at least some FRBs. If the intrinsic mechanism originates from compact emission regions on scales of $\sim$ kms or AUs, then FRBs are effectively a point source on cosmological scales of $\sim$ Gpcs. Then, FRBs can probe the propagation path from the emitter to an observer at a precise scale \cite{Kumar2024}. One common propagation effect observed for a point source or ``unresolved'' emitter is scintillation. Scintillation is the modulation in flux that occurs due to the coherent interference of radio waves generating a diffraction pattern in frequency \cite{Salpeter1967}. In \textcite{Nimmo2025}, the detection of two scintillation screens for FRB 20221022A provides evidence that FRBs can act like point sources along the majority of a sight line. There are thousands of bright FRBs per day \cite{CHIMEFRB2021} and as these bursts originate from extragalactic distances, FRBs can act as a point source probe on cosmological scales. Along the propagation path, FRBs could encounter dark compact objects, which may be primordial black holes \cite{Munoz2016,Eichler2017,Katz2020}, stars \cite{Connor2023}, or rogue planets \cite{Jow2020}. If such an encounter was able to preserve phase coherence, one could unambiguously detect these unique encounters \cite{Kader2022,Leung2022}.

Throughout this paper, we refer to the phase coherence of the electric field. This phase coherence, sometimes referred to as the mutual coherence of the electric field \cite{Schneider1985}, can exist between two paths of propagation for a compact source. This is the same property that allows two radio antennas to act as an interferometer, separated by a baseline distance, to coherently interfere with each other to image the sky. We have previously developed a pipeline to detect the existence of this phenomenon for a lens that does not depend on the frequency of the observation\cite{Kader2022} and extended the formalism to find one that does \cite{Kader2024}. In this work, we will further apply these techniques to determine the possibility of coherent lensing for an FRB. For an FRB, there are two relevant types of physical lenses: a gravitational lens \cite{Munoz2016} and a plasma lens \cite{Clegg1998}. A plasma lens is created when there exists an inhomogeneity in the electron density along the line of sight to an FRB \cite{Cordes2017}. A plasma lens is an angularly structured distribution of the change in electron density along the line of sight. Observationally, this is a separate phenomenon from the average electron column density that produces the observed dispersion measure (DM) of radio bursts and the stochastic inhomogeneities that produce scattering and scintillation \cite{Salpeter1967, Rickett1990} but the underlying propagation physics for all these effects is the same.

The concept of lensing due to discrete plasma structures is sometimes referred to as extreme scattering events, which was initially observed through observations of quasars \cite{Fiedler1987}. Observations of some pulsars have also found that discrete plasma structures can alter the burst morphology and observed properties. Plasma lensing of pulsars has been observed to produce images (or ``echoes") of the emission \cite{Backer2000, Main2018, SerafinNadeau2025}. FRBs are expected to be compact emission sources like pulsars, such that the propagation effects should produce similar observational signatures. The environment around an FRB is thought to contain dense, magnetized plasma \cite{Masui2015,Michilli2018}, which itself can provide the conditions for lensing to occur. It is possible that some morphological properties of the FRB bursts can arise from plasma lensing. For example, some FRBs are observed to be quite narrowband in frequency, which might be caused by a near-caustic magnification of the intrinsic burst \cite{Kumar2024a}. This interpretation, however, implies that some broadband FRBs must also be observed with a brighter narrowband feature near a caustic. The observation of some FRBs does suggest the changes in morphology and dispersion measure (DM) may arise from plasma lensing \cite{Cordes2017,Chen2024,Platts2021}. We are therefore motivated to search for signatures that may indicate whether the complex morphology of the FRB is the emission mechanism itself or propagation through the complex environment surrounding the burst.

In this paper, we report on FRB 20220413B observed with the Canadian Hydrogen Intensity Mapping Experiment (CHIME). It shows evidence for partially phase coherent propagation and morphological similarity to a near-caustic plasma lensed FRB. In section \ref{sec:data}, we introduce the data product and summarize the observed burst properties. In section \ref{sec:morphfit}, we perform a morphological fit to the distinct branching structure observed in the intensity burst profile, assuming a plasma lensing scenario. In section \ref{sec:tcorr}, we analyze the voltage data to search for phase coherence of the electric field, while in section \ref{sec:fcorr}, we analyze the intensity spectra of the burst components and fit for a scintillation bandwidth. Finally, in section \ref{sec:scenarios}, we simulate possible scenarios that can produce the observed phase correlation signatures and discuss the implications.

\section{Data and Morphology}\label{sec:data}
CHIME \cite{CHIME2022} is a radio telescope located near Penticton, British Columbia, Canada. It is a drift scan telescope consisting of 4 cylindrical reflectors with no moving parts. Each cylinder consists of 256 dual-polarization antenna feeds. CHIME is able to record complex-valued voltage data across 1024 frequency channels between 400 and 800 MHz, sampled every 2.56 $\mu$s. This channelized complex-valued voltage data will be referred to as baseband data hereafter. The CHIME/FRB search backend \cite{CHIMEFRB2018} can record the baseband data for an FRB detection. This dataset per event typically provides a search time for a coherent lensing search at CHIME up to $\sim 100$ ms \cite{Kader2022}. The baseband analysis pipeline \cite{Michilli2021} converts and processes the baseband data from all feeds into a single data product that is beamformed to the location of the FRB and coherently dedispersed. We use this baseband data product for all the analyses presented in this paper.

\begin{figure}[!htb]
    \centering
    \includegraphics[width=0.35\linewidth]{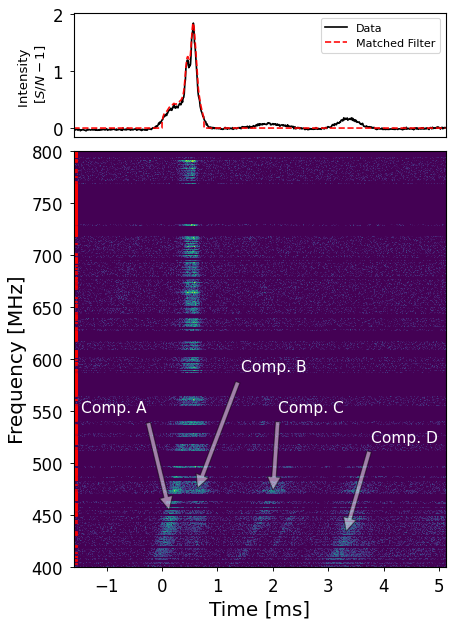}
    \caption{The intensity profile of FRB 20220413B at 2.56 $\mu$s resolution dedispersed to the S/N maximizing DM is shown in the bottom panel, while the top panel contains the frequency-averaged intensity profile in units of S/N. The four burst components used in the presented analyses are labeled from left to right as A, B, C, and D. The burst contains many components featuring an upward drift structure and a possible bifurcation of the burst into components A and B around $\sim 550$ MHz. Masked frequency channels are marked with red.}
    \label{fig:burst_morph}
\end{figure}

In \textcite{Faber2024}, a morphological analysis was done on the baseband burst profile for several FRBs observed with CHIME. One particular FRB, FRB 20220413B, had a morphology that might be consistent with plasma lensing. That is to say, the spectro-temporal profile of this burst could be generated through the process of an FRB propagating through an inhomogeneous plasma acting as a lens. Figure \ref{fig:burst_morph} shows the total power of the burst summed over both polarizations. We label the components of this burst from left to right in time, labeled from A to D. There is an apparent bifurcation $\sim$550 MHz in the broadband component of this burst, of which we have labeled the split components as A and B. Components C and D are the brightest components left of the broadband structure, respectively. There exists fainter sub-components with lower S/N ratios than components A, B, C, and D, which we do not consider for the analyses conducted in this paper. If plasma lensing is the phenomenon generating these morphologies, then the fainter components would be manifestations of the same phenomenon. If this is not due to plasma lensing, then they may be related to the intrinsic emission process.

\begin{table}[h!]
\centering
\begin{tabular}{c|c|c|c|c} 
 \hline
  R.A. & Dec & $\mathrm{DM}_{A,D}$ & $\mathrm{DM}_{B}$ & $\mathrm{DM}_C$ \\
  ~(J2000) & (J2000) & [$\mathrm{pc}~\mathrm{cm}^{-3}$] & [$\mathrm{pc}~\mathrm{cm}^{-3}$] & [$\mathrm{pc}~\mathrm{cm}^{-3}$] \\  
 \hline
 $16^{h}49^{m}03^{s}$ & $+66^{\circ}58'56''$ & 115.673  & 115.723 & 115.613 \\
 \hline
\end{tabular}
\caption{Relevant burst properties for FRB 20220413B, obtained from \citet{Faber2024}.}
\label{tab:burstp}
\end{table}

In \citet{Faber2024}, the morphology analysis of this burst finds a frequency-time scaling relation, $t \propto f^{\alpha}$, for component A to be $\alpha=-3.9(5)$. If, instead, the bifurcation is real, then there exists a branching point at a critical frequency and a DM following the typical dispersive $\alpha=-2$ scaling \cite{Faber2024}. For determining burst properties and dedispersing the data, we consider the DM values in this formalism. The burst properties are summarized in table \ref{tab:burstp}, including the different DMs for the different components. The main DM value for this burst is obtained as the S/N maximizing DM, given by $\mathrm{DM}_{B}$.

This bifurcation feature is highly suggestive of a lensing morphology. If we assume lensing is responsible for this feature, then it must be a near-caustic lensing interaction. A caustic occurs when images merge and can produce a significant increase in the flux at the merging point. The frequency dependence of a plasma lens can allow for the merging point to be directly observed at a critical frequency. The branching point of the bifurcation present in FRB 20220413B would be the critical frequency of a near-caustic plasma lens. In the next section, we constrain and fit a plasma lens model to this burst morphology under this assumption. 

\section{Fitting for a lensing Morphology}\label{sec:morphfit}
In this section, we fit a plasma lensing model to the burst morphology. The morphology of FRB 20220413B might be indicative of plasma lensing. In particular, we are focused on the branching morphology that splits into components A and B. Lensing is a difficult phenomenon to directly fit and constrain due to a non-linear dependence of the fit parameter space and the number of images generated. This section will present an overview of how we can fit a plasma lens model given the constraint that we observe the caustic point.

Before we describe how plasma lensing can generate the observed burst morphology, we briefly introduce the formalism here. For a more detailed description, we refer the reader to \cite{Kader2024,Feldbrugge2019,Jow2023}. For a single lens, the Fermat potential describing the time delay potential is given by, 
\begin{equation}\label{eq:ferm_pot}
    T  = \frac{\mu_g}{2} (x_i-y_i)^2 + \mu_l \Phi(x_i).
\end{equation}
Equation \ref{eq:ferm_pot} is described by the variables, $x_i$ which is the unitless parameterization of the 2D lensing plane, $y_i$ which is the unitless parameterization of the location of a point source on the source plane, $\mu_g$ is the geometric parameter, $\mu_l$ is the lensing parameter, and $\Phi(x_i)$ is the lensing model. In physical parameters, we have $\mu_g= (1+z_l)D_{ol} D_{os}\theta_c^2 c^{-1} D_{ls}^{-1}$ where $z_l$ is the redshift of the lens, $D_{ol}$ , $ D_{os}$ , $D_{ls}$ are the angular diameter distances, and $\theta_c^2$ is some characteristic lens scale. For the lensing parameter, a plasma lens \cite{Clegg1998} will have $\mu_l= k_{\mathrm{DM}} \mathrm{DM}_l f^{-2}$ where $k_\mathrm{DM}=4.149 \times 10^{15}$ $\mathrm{s}~{\mathrm{Hz}}^{2}~\mathrm{{pc}}^{-1}~ {\mathrm{cm}}^{3}$, the DM constant, $f$ is the observing frequency, and $\mathrm{DM}_l$ is the DM of the lens. This function describes all possible paths of propagation for a point source at $y_i$ to a point observer located at the origin. The Kirchoff-Fresnel diffraction integral (KDI), $\mathbf{E} = \int d \mathbf{x} \mathbf{E}_s e^{i2\pi f T} $, provides the observed electric field after propagation through a lens. This is a highly oscillatory integral that is difficult to evaluate computationally outside of specific conditions \cite{Feldbrugge2019,Grillo2018}. In the limit where the distances are large compared to the wavelength, the semi-classical approximation will largely hold, such that images are given by the stationary points of this integral \cite{Nakamura1999}. 

In the semi-classical approximation, images form where $\partial_i T = 0 $. Suppose a lens generates $l$ images. Then, for these images, $\mathbf{x}^l$, the time delay relative to the unperturbed path of an image is $\tau_l = T(\mathbf{x})$. The Hessian of the Fermat potential, $\partial_{ij} T$, provides the local curvature around an image through its eigenvalues determined at that point, $\lambda_{\pm}(\mathbf{x}^l)$. The image magnification is given as $\varepsilon_l = \left(\lambda_+(\mathbf{x}^l) \lambda_-(\mathbf{x}^l)  \right)^{-1/2}$. If either eigenvalue is 0, then the image point lies on a caustic. For lensing occurring through a single lens and where the total extent of the images is significantly smaller than the beam width, the propagation transfer function of the lens is given as
\begin{equation}
    H(f) = \sum_l \varepsilon_l e^{\i 2\pi f \tau_l } .
\end{equation}
In \textcite{Kader2024}, a simulation toolset was developed to model the coherent signatures and determine this propagation transfer function. We use this toolset to model and fit for a lensing system by directly fitting for $\mu_g$, $\mu_l$, and $y_1$, assuming a 1D lens model for simplicity and faster computation times. We use a simple plasma lens model given by a 1D gaussian \cite{Clegg1998},
\begin{equation}
    \Phi(x_i) = \exp\left(-\frac{1}{2}x^2_1\right).
\end{equation}
With this lens model, there exists three parameters that determine the Fermat potential $\mu_g$, $\mu_l$, and $y_1$. With a set of these parameters, we generate a set of images, $l$, for each of the 1024 frequency channels, where every image has a delay $\tau^l$ and a magnification $\varepsilon^l$.  Using these observables, we can create image copies of an intrinsic burst function with a relative delay and magnification applied. These are relative as we reference them to the delay and magnification of the first image at $800$ MHz to mimic the burst as observed after it has been dedispersed by the system. The total sum of images will model the intensity profile of the FRB. 

We model the intrinsic burst profile to be a Gaussian function with no spectral index in observed flux. The full model function is given by,
\begin{equation}
    M(f,t) = \sum_{l} A |\varepsilon^l(f)|^2 \exp\left(-\frac{(t-t_0-\tau^l(f)-\tau_{\Delta\mathrm{DM}}(f))^2}{2\sigma^2}\right),
\end{equation}
where $A$ is the amplitude of the burst in every channel, $\sigma$ is the width of the burst, $t_0$ is the arrival time, and $\Delta \mathrm{DM}$ is a DM error parameter, separate from $\mathrm{DM}_{lens}$, to account for discrepancies between the S/N maximized DM (found in table \ref{tab:modelfit}) and the constant DM term needed to morphologically fit to model plasma lens to the data.

\begin{figure}[!htb]
    \centering
    \includegraphics[width=0.9\linewidth]{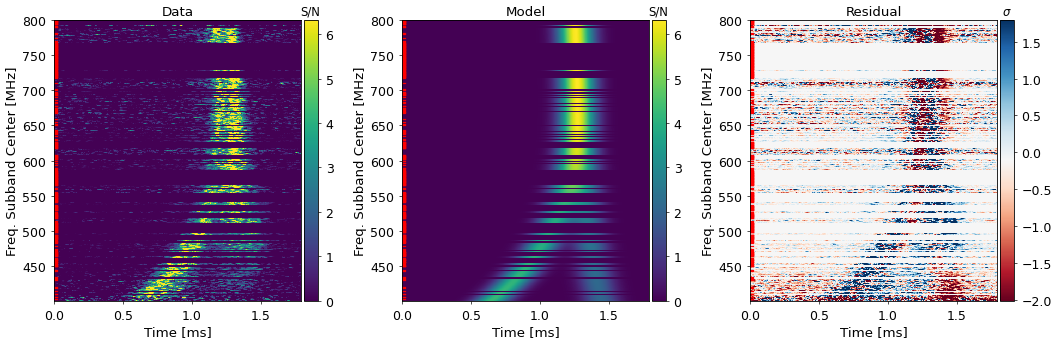}
    \caption{The fitted lensing model to FRB 20220413B. A 1D Gaussian plasma lens near a cusp caustic is able to replicate the observed frequency-time morphology. Key discrepancies are seen in the observed flux at frequencies $<$ 500 MHz and an apparent dual burst component at frequencies $>$ 600 MHz. This may be due to the emission process or a more complex plasma lens distribution. Masked channels are indicated in red.  }
    \label{fig:ftime_fit}
\end{figure}

We note that fitting for a lens model by directly obtaining the Fermat potential is rather difficult for reasons we will discuss. Initial attempts at minimizing the typical residual function, $ R(f,t) = I(f,t) - M(f,t)$, found little success in convergence due to the non-linear dependence between lens parameters and the resulting model morphology. The lens parameters $\mu_g$, $\mu_l$, and $y_1$ can generate anywhere between 1 to 3 images, which implies the parameter space is discontinuous by nature. To further complicate the scenario, we consider the bifurcation in this burst to be from images forming near a caustic. Caustics arise when one or more of the eigenvalues of the Hessian of the Fermat potential are zero, i.e. $\lambda_{\pm}(\mathbf{x}^l)=0$. Images merge, changing the total number of images, which causes discontinuities in the residual function. A discontinuous parameter space falls under the regime of catastrophe theory \cite{Saunders1980,Berry1980}. The 1D Gaussian plasma lens is topologically similar to a cusp catastrophe \cite{Clegg1998,Feldbrugge2019,Jow2023} such that we can approximate the image generations based on a cubic polynomial (see appendix \ref{sec:cuspfit}). We directly fit the caustic cusp morphology to obtain the physical parameters associated with it. The critical merging point for the 1D Gaussian lens is approximately given by the solution to the set of equations (see appendix \ref{sec:cuspfit} for the derivation),
\begin{equation}
\begin{split}
     0  &=  12- 12\alpha^2  -y^2 \\
     0  &= -y \left(72 + 144 \alpha^2 +216\alpha +2y^2  \right) ,\\
\end{split}
\end{equation}
where $\alpha = \mu_l \mu_g^{-1}$ and $y$ is the unitless angular source offset. This set of equations comes from the solutions to the bifurcation set for a cubic cusp potential (see eq. \ref{eq:cuspparams} and eq. \ref{eq:cuspparamscrit} in appendix \ref{sec:cuspfit}). This provides a relation between the lens parameters, such that we can determine a critical lens strength for a given source offset, $\alpha_{crit}(y)$. Furthermore, when both $\alpha$ and $y$ are solutions to the critical equations, then these parameters correspond to the caustic parameters and the point of maximum magnification in the frequency-time domain (see appendix \ref{sec:cuspfit}). Here we note $\alpha$ has a dependence on both the frequency and the DM of the lens following $\alpha \propto \mathrm{DM}_{lens} f^{-2}$ such that we may bound $f$ between 400$-$800 MHz, and fit the lens model using parameters $f_{crit}$, $y$, and $\mathrm{DM}_{lens} $. 
Having a fit near a caustic using a semi-classical approximation inherently introduces singularities in the magnification. We can account for this by normalizing our model to the total power in time and altering our residual function to be of the form,
\begin{equation}
    R(f,t) =\frac{ I(f,t)}{\sum_tI(f,t)} - \frac{ M(f,t)}{\sum_tM(f,t)} . 
\end{equation}
Though we know the magnification at caustics will be wrong, the frequency-time relation will be correct, and therefore, this residual function ensures we are fitting for the morphology. Additionally, we are able to account for variations due to the unknown, intrinsic spectrum of the FRB by computing this residual function separately for each frequency channel. The fit is shown in figure \ref{fig:ftime_fit}. We find the fit converges to a $\chi^2_{red}=2.9$ with fit parameters found in table \ref{tab:modelfit}.

\begin{table}[!htb]
    \centering
    \begin{tabular}{c|c|c|c|c|c|c}
      \hline     
     $A$   & $\sigma$ & $t_0$  & $f_{crit}$ & $\mathrm{DM}_{lens}$  & $y$ & $\Delta \mathrm{DM}$ \\
     ~[S/N]  & [$\mu$s] & [$\mu$s] &  [MHz] &  [pc $\mathrm{cm}^{-3}$] & [Arb. Units] & [pc $\mathrm{cm}^{-3}$]  \\ 
    \hline
     $4.45 \pm 0.01$  &  $82.2 \pm 0.2$  & $8167.0 \pm 0.4$ & $727.73 \pm 0.03 $ & $0.0888 \pm 0.0003$ & $0.0083 \pm 0.0005$ & $0.01022 \pm 0.00009$ \\
      \hline
    \end{tabular}
    \caption{The morphological fit parameters found for FRB 20220413B using a 1D Gaussian plasma lens model.}
    \label{tab:modelfit}
\end{table}

We note the key parameters to describe the lens model are $f_{crit}$, $y$, and $\mathrm{DM}_{lens} $. With these parameters, we can determine $\alpha_{crit}$ and obtain the frequency-independent, geometric lens parameter, $\mu_{g}\sim 0.7$ ms. Additionally, it is important to note that when we fit a lensing model, we are not seeking to find the model that explains the full propagation path. Instead, we are determining whether there exists a model that can explain the observed signatures. Given this consideration, we can see in figure \ref{fig:ftime_fit} that the branching morphology can be replicated by this simple plasma lens model. It is not exact, and there are significant discrepancies that should be highlighted here. There are two components in the data in the upper part of the frequency band ($>$600 MHz) that are not modeled by our simple plasma lens model. The lack of signal in the right-most branch of the bifurcation of the burst in the data compared to the model is important. Both distinctions may be caused by either the FRB emission mechanism or a more complex lensing model. 

If the observed components are image copies of the initial electric field rather than some unknown intrinsic process, then there will exist phase coherence between the bursts. We can search for phase coherence by time-lag correlating the burst components. This would be a confirmation of the lensing scenario. This search is conducted using the complex-valued voltage data and provides information about temporal correlations. 

\section{Time-lag Correlation}\label{sec:tcorr}
We consider the scenario of coherent lensing where a lens creates multiple image copies of the electric field, where the phase of the wavefront and its relation between images is preserved, i.e. coherent. This is analogous to the baseline delay between two radio interferometers that observe the same electric field, which will coherently correlate in phase. In previous work \cite{Kader2022}, we developed a pipeline to identify coherent correlation signatures that arise from gravitationally lensed FRBs. Plasma lensing is different from a gravitational lens as it is chromatic due to the frequency dependence of the group velocity as the electric field propagates through the plasma \cite{Clegg1998}. To identify coherent signals that are dependent on frequency and produced by plasma lensing rather than gravitational lensing, we modify the correlation algorithm from our previous work. The full details of identifying coherence with a frequency-dependent time-lag correlation can be found in \citet{Kader2024}. The time-lag correlation per frequency channel is given by,
\begin{equation}\label{eq:tcorr}
    <C(f,\hat{t})> = \frac{\int dt V(f,t) V^*(f,t-\hat{t}) W^2(f,t) }{\sqrt{\int dt |V(f,t)|^2 |V(f,t-\hat{t})|^2 W^4(f,t)}},
\end{equation}
where $V(f,t)$ is the baseband data and $W^2(f,t)$ is the matched filter and is proportional to the frequency summed intensity of the main burst, $W^2(f,t) \propto \int df |V(f,t)|^2$. It should be noted here that this is not the optimal filter for the burst, as frequency information is averaged together, whereas an optimal search for plasma lensed images might include the intensity profile as a function of frequency. Instead, this filter provides equal weight to all frequencies so as not to impose assumptions about the lens model or the intrinsic FRB spectrum. This will identify coherent signals, if the phase delay is constant across a frequency channel, as a delta function response \cite{Kader2024}. A coherent phase is identified per channel center $f_c$ by requiring a constant phase over a channel bandwidth $\Delta f$, where coherence can be measured if $ e^{\i2 \pi( f_c +\Delta f) t} e^{\approx \i2 \pi f_ct }$. At the native baseband resolution, the channel bandwidth is $\Delta f = 390.625$ kHz. Such a signal would indicate the electric field of the filtered burst has been replicated, which is expected for a coherent lensing event. 

To evaluate the extent to which correlated signals are derived from phase correlations, we create a mock realization of our dataset by randomizing the phase for one of the baseband inputs while preserving power and performing the same time-lag correlation. Formally, this is defined as $V_{mock}(f,t)=|V(f,t)|e^{\i \phi_{mock}(f,t)}$ where $\phi_{mock}(f,t)$ is a uniform random variable sampled between $[0,2\pi)$. We perform the same procedure with this mock dataset to validate any signatures in the dataset. Any signature present in the time-lag correlation of the baseband data that is inconsistent with the mock is considered a candidate coherent lensing event.

\begin{figure}[!htb]
    \centering
    \includegraphics[width=0.7\linewidth]{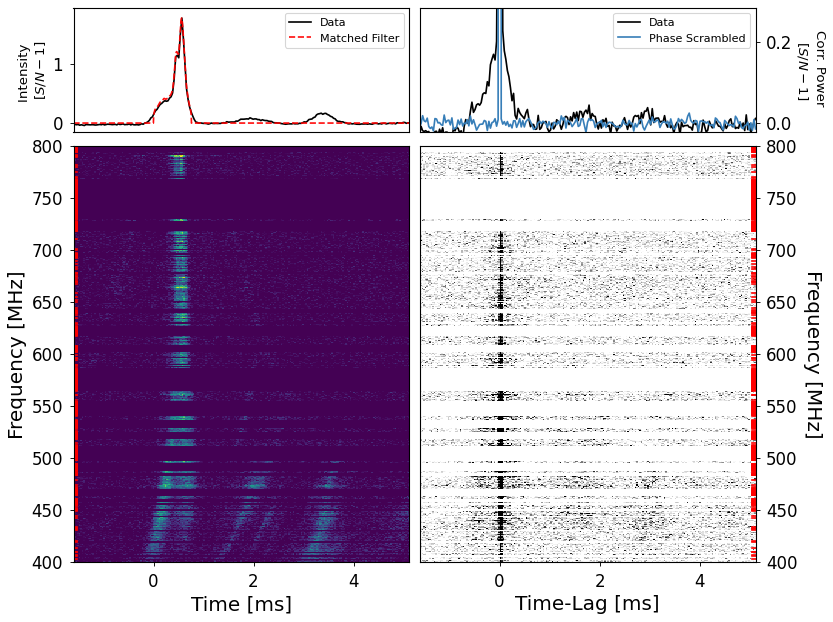}
    \caption{ The power of the time-lag correlation for FRB 20220413B is shown in the left panels, while the intensity of the burst is shown in the right panels. The burst shows excess power in the time-lag around $\sim 1-3$ ms that matches the location of time-separated bursts. The frequency average power is shown in the top panels. The intensity profile additionally shows the matched filter (dashed red) while the time-lag power profile shows the measured response (black) and the phase scrambled mock response (blue). Masked frequency channels are marked with red.}
    \label{fig:burst_ftimecorr}
\end{figure}

We present the time-lag correlation of FRB 20220413B in figure \ref{fig:burst_ftimecorr}. The intensity profile of the burst is shown in the left plots of figure \ref{fig:burst_ftimecorr}. The power of the matched filtered time-lag correlation of components A and B is shown in the right plots, i.e. $|C(f,\hat{t})|^2$. Both top panels are the frequency-averaged power in units of excess S/N relative to the noise expectations, i.e. $S/N - 1$. The mock response, represented by the blue line in the top right panel, is compared to the measured correlation shown by the black line. The excess power found in the correlation response of the data compared to the mock response highlights that this signature is real and is present at various time-lags. The excess power is found at time-lags directly corresponding to the separation of components in FRB 20220413B. This is evidence that there is a correlation signature present for this burst. 

In the case of an image produced by a gravitational lens, the time delay of the signal is found independent of frequency. For a plasma lens, images will have a difference in DM, changing the phase delay to be dependent on frequency \cite{Kader2024}. In our previous search \cite{Kader2022}, we optimized our algorithm to find achromatic images. For this method, we require the frequency information and therefore search for evidence of an image by further performing an incoherent DM search in this domain.  

\begin{figure}[!htb]
    \centering
    \includegraphics[width=0.7\linewidth]{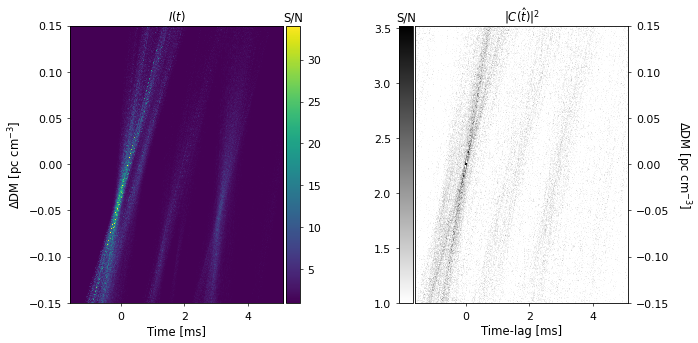}
    \caption{ A DM search was applied to the intensity profile (left) and time-lag correlation (right) of FRB 20220413B over 400 - 500 MHz. A coherent DM phase delay will create a delta function response at a specific DM and time-lag if one exists. We find an increase in S/N corresponding to time-separated burst components similar to the DM search of the intensity profile. We do not find an exact DM phase difference between components in the DM search of the time-lag correlation. The zero-lag and $\pm2.56$ $\mu$s, $\pm5.12$ $\mu$s lags were masked in this search. These lags are known to correlate due to the CHIME channelization process. }
    \label{fig:burst_dmsearch}
\end{figure}

There are known systematic delays at 1 and 2 integer multiples of $\pm2.56$ $\mu$s corresponding to the polyphase filterbank used in the signal processing chain of CHIME \cite{CHIME2022,Kader2022}. Therefore, we ensure these delays are masked for the DM search in the time-lag domain. We will further impose the search to be over 100 MHz from 400 - 500 MHz to account for the complex burst morphology. Figure \ref{fig:burst_dmsearch} shows the DM search in the time-lag domain in the right panel and the DM search in the time domain in the left panel. We find no evidence to suggest that there exists a coherent phase delay across 400 - 500 MHz for FRB 20220413B. There is an evident increase in the signal-to-noise ratio corresponding to the time separation between different components, seen in both the time and time-lag domains. This highlights that the increase in the variance is correlated with the burst components. However, we find no evidence to suggest there is an increase in the S/N at a specific DM and time-lag value. If there did exist a phase coherent signal, the S/N should increase relative to the background, which can be seen to occur in the right panel of figure \ref{fig:burst_dmsearch} near DM $=0$ pc $\mathrm{cm}^{-3}$ and $0$ ms lag but not for the other components at lags $\sim-0.5, 1.5,$ and $\sim 3$ ms. 

Based on this, we find no statistical evidence for a coherent phase correlation in the time-lag correlation. We do see an excess in the power of the time-lag correlation relative to the expectations for no phase signature. If no coherent phase relation exists, then there are no coherent multipath propagations that exist from the FRB source to the observer. With a signature present only in power, it may be possible that there is coherent multipath propagation for this burst, but from an unresolved scattering screen. To investigate this, we perform a scintillation analysis on the components of this burst. We will correlate the intensity spectra of all components with themselves and each other to search for frequency correlations. 

\section{Frequency-Lag Correlation}\label{sec:fcorr}
\begin{figure}[!htb]
    \centering
    \includegraphics[width=0.7\linewidth]{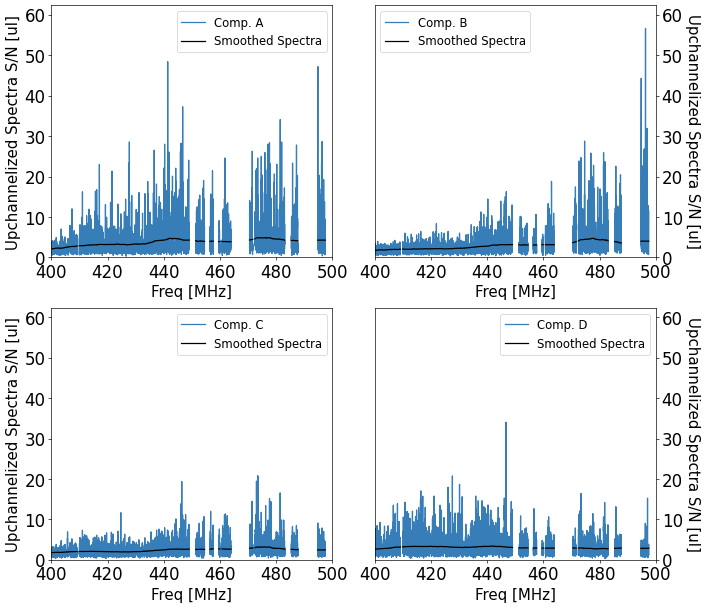}
    \caption{The upchannelized spectra of the four components that compose FRB 20220413B between 400 - 500 MHz are shown in blue, while the smoothed spectra are shown in black. There is a large variation in S/N and spectral structure between the burst components.}
    \label{fig:burst_fspecs}
\end{figure}

The detection of excess S/N in the time-lag correlation of the burst presents the possibility that there exists coherent multipath propagation for this FRB, including plasma lensing. To further understand the propagation effects that are affecting this burst, we search for signatures of scintillation present in the spectra of this burst. Scintillation is the frequency-dependent modulation of the spectra of the burst originating from the multipath propagation through inhomogeneous cold plasma along the line of sight \cite{Rickett1977}. Scintillation is a coherent process that requires an unresolved source to generate these modulations. Therefore, if this burst has a common scintillation present in all components, then it is likely that there is an unresolved source and a common multipath propagation present for these separate components. A common propagation would generate the correlation response observed in figure \ref{fig:burst_ftimecorr}.

\begin{figure}[!htb]
    \centering
    \includegraphics[width=0.7\linewidth]{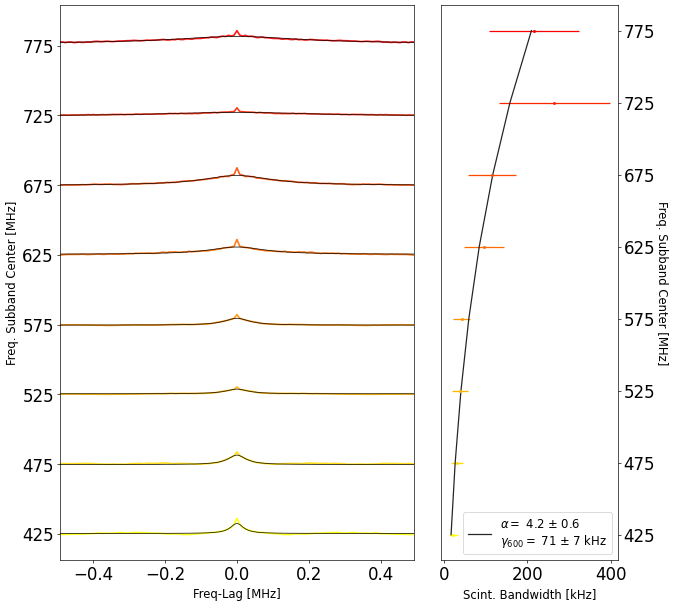}
    \caption{ The intensity spectra component B of the FRB is split into 8 sub-bands, each 50 MHz in width. A Lorentzian function is fit to the frequency-lag correlation for each sub-band to obtain the scintillation bandwidth, shown on the left panel. The right panel shows the fit to the scintillation scaling as a function of frequency. The best fit has a scaling index of $\alpha= 4.2\pm0.6$ and a bandwidth of $\gamma_{scint}=71\pm 7$ kHz at 600 MHz, consistent with expectations for scintillation.}
    \label{fig:burst_scintscaling}
\end{figure}

The lack of any statistically significant response in the time-lag correlation suggests things are not fully coherent. The excess in coherent variance suggests that there may exist a scattering screen causing diffractive scintillation. To investigate this, we can perform a frequency-lag correlation of the intensity spectra of the burst components. For the analysis, we follow the method in \textcite{Nimmo2025} with modifications to allow for the cross-correlation of components. We note the general method and distinctions as follows. First, we upchannelized the baseband data, using Fast Fourier transforms to convert the native baseband data with a resolution of $2.56$ $\mu$s and $390.625$ kHz to an upchannelized dataset with a resolution of  $50\times2.56= 128$ $\mu$s and 7.8125 kHz. We use the upchannelized dataset dedispersed to the corresponding DM of the component (table \ref{tab:burstp}) and individually select the burst components and obtain the average intensity spectra $I(f)$ for each component, shown in blue in figure \ref{fig:burst_fspecs}. We then obtain a smoothed spectrum $\bar{I}(f)$ by taking a moving average of $I(f)$ over a scale of $10$ MHz, shown in black in figure \ref{fig:burst_fspecs}. Systematic fluctuations from the beam response of CHIME are present at these scales \cite{Nimmo2025,CHIME2022}, so we are focused on modulations occurring at smaller scales. Additionally, the smoothed spectra, in contrast to a power law fit, should help account for the complex and narrowband intensity variations seen for these components. Finally, we make a modification to the general frequency-lag correlation \cite{Nimmo2025,Macquart2019,Sammons2023} by generalizing it to be the cross-correlation between two components rather than the auto-correlation between one component. The frequency-lag correlation of the intensity spectra between two burst components is then given as,
\begin{equation}\label{eq:fcorr}
    C(\hat{f}) = \frac{\int df (I_1(f) - \bar{I_1}(f)) (I_2(f-\hat{f}) - \bar{I_2}(f))}{\int df\bar{I}_1(f)  \int df \bar{I}_2(f)} .
\end{equation}
Only component B features a broadband spectrum, so we subdivide that spectrum into 8 sub-bands, where each sub-band has a width of 50 MHz. For each frequency-lag correlation, we fit a Lorentzian function \cite{Nimmo2025},
\begin{equation}
    C(\hat{f}) = \frac{m^2}{1+\frac{\hat{f}^2}{\gamma_{scint}^2}} + c,
\end{equation}
where $m$ is the modulation index, $c$ is a mean offset, and $\gamma_{scint}$ is the scintillation bandwidth, which scales with frequency,
\begin{equation}
    \gamma_{scint} \propto f^{\alpha},
\end{equation}
with a scaling index $\alpha$. A scaling index with $\alpha=4$ or $\alpha=4.4$ is indicative of scintillation and scattering of a point source \cite{Lee1975}. Figure \ref{fig:burst_scintscaling} shows the fit for each scintillation bandwidth in the left panel and the scaling fit in the right panel. We fit a scintillation bandwidth of $71 \pm 7$ kHz at $600$ MHz and a frequency scaling of $4.2 \pm 0.6$ with a reduced $\chi^2=0.28$. We find this scaling index to be consistent with expectations for scattering, both 4 and 4.4. 

With a measure of the scintillation bandwidth for burst component B, we know scintillation is present for this burst. The question that remains is whether this is the reason we observe the correlations present in the time-lag correlation of the burst (section \ref{sec:tcorr}). That is to say, diffractive scintillation implies coherent phase propagation which requires a source to be unresolved \cite{Jow2022,Rickett1977,Narayan1992} and if we observe the same scintillation in all the burst components then this entire burst is likely unresolved \cite{Sammons2023} to a scattering medium which would create the correlations present in the time-lag correlation due to a common coherent multipath propagation. 

\begin{figure}[!htb]
    \centering
    \includegraphics[width=0.7\linewidth]{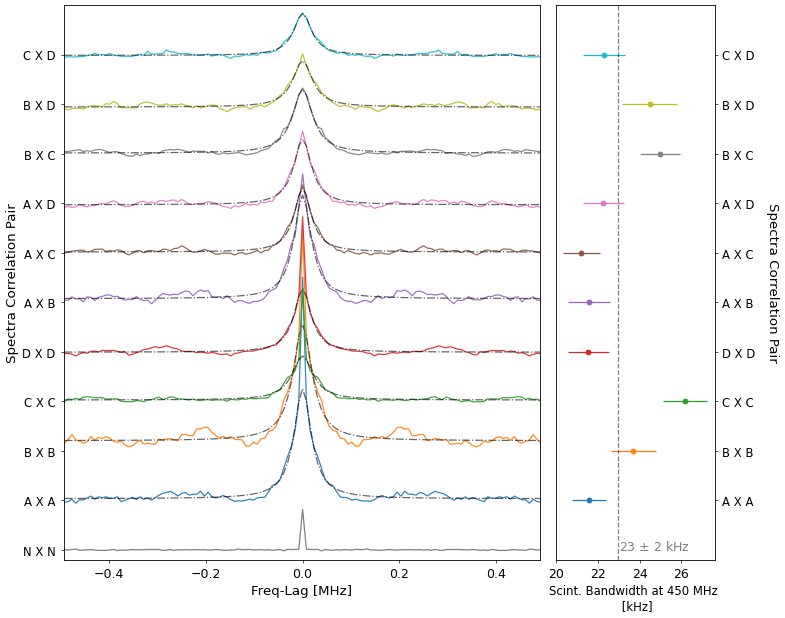}
    \caption{The frequency-lag correlation for all burst components (A, B, C, D) and the noise expectation (N) is shown in the left panel. The frequency-lag correlations are computed between common pairs (e.g. A $\times$ A) and cross pairs (e.g. A $\times$ B). The fitted scintillation bandwidth distribution (grey dash-dot line in left panel) for each pair is shown on the right panel. There is a similar bandwidth measured for both the common and cross pairs, indicating all components have the same scintillation present. The average scintillation bandwidth at 450 MHz for all components is $23\pm2$ kHz. }
    \label{fig:burst_scintcrosscorr}
\end{figure}

We correlate, in frequency, the intensity spectra of each burst component (A, B, C, and D) with itself and with each other. The frequency-lag correlation for all pairs is shown in the left panel of figure \ref{fig:burst_scintcrosscorr}. We, additionally, present the correlation of an off-pulse region containing only noise and no burst components  (labeled N $\times$ N) to show that equation \ref{eq:fcorr} provides a null response when no oscillations in the spectra are present. For all burst component pairs, there is a visible similarity in the width of the central component and subsequent oscillations at frequency-lags $> 0.2$ MHz. Additionally, the cross-correlation response is seen to peak at a frequency-lag near 0. These features are all strong indicators that the scintillation pattern is the same for all burst components. To quantify this, we fit the Lorentzian function to obtain the modulation index and scintillation bandwidth at 450 MHz for all frequency-lag correlation pairs. The fitted scintillation bandwidth at 450 MHz is shown in the right panel of figure \ref{fig:burst_scintcrosscorr}. The scintillation bandwidth appears to be largely consistent for all pairs. We find an average scintillation bandwidth for all pairs of $23 \pm 2$ kHz at 450 MHz, consistent with the fitted bandwidth for component B, $\gamma_{scint}= 21$ kHz scaled to $450$ MHz. The expected scintillation bandwidth from the NE2001 survey \cite{Cordes2002,Ocker2024} at 450 MHz is $31$ kHz. The discrepancy between the expected and measured average value is $\sim 26 \%$. This is relatively consistent given the expected errors of the NE2001 model \cite{Cordes2002} and is evidence that the scattering screen generating the scintillation originates from the Milky Way rather than in the host galaxy. 

We note there is no visible evidence that the frequency-lag correlation for any of the cross pairs contains a zero-lag peak in addition to the Lorentzian response. That is to say, for the frequency-lag correlation between the same pair, e.g. A $\times$ A, there is a zero-lag response as the intrinsic spectrum is the same and the scintillation pattern is the same. If the cross-pair correlation, e.g. A $\times$ B, was the correlation between images, where A is B intrinsically, then a fully coherent cross-pair correlation would result in a response like the auto-pair correlation response. There would be an additional contribution at zero-lag, creating a delta response separate from the Lorentzian. We do not see any strong evidence for this, which is consistent with results from the time-lag correlation of the burst components.

The modulation index for these spectra may further indicate whether the emission is resolved by scattering near the host \cite{Main2021}, or if the Milky Way scattering screen has resolved the host screen \cite{Nimmo2025,Sammons2023,Kumar2024,PradeepE.T.2025}, or if plasma lensing has occurred and the plasma lens is resolved by the host scattering screen. In order to assess and interpret the modulation index, CHIME systematics\cite{Nimmo2025} must be better understood. We find this beyond the scope of this paper and leave this investigation for future work. We note this systematic uncertainty does not affect the interpretation of the results presented in this section, as an incorrect measure of the modulation index cannot cause the scintillation bandwidth to be consistent across all correlation pairs.

The analysis of the intensity spectra for all burst components finds scintillation is present for this burst, and there exists a common scintillation pattern present for all components. This implies there is a common propagation path through a scattering medium that is creating the scintillation pattern common to all burst components. In the previous section, we found a time-lag correlation signature present only in the power of the time-lag correlation and not the phase for the time-separated burst components. With a confirmation that all components have a common scintillation bandwidth, the likely scenario is that the correlation in power arises from the coherent multi-path propagation that generates the scintillation. Furthermore, the measured scintillation bandwidths are consistent with the expectations from scattering through the Milky Way, indicating that the scintillation is contributed by a Galactic screen. With no observed coherent phase between components, we do not find any evidence for fully coherent lensing. Thus, while we find coherent path propagation has occurred along the path of propagation, the question remains whether this FRB is a lensed FRB that has experienced phase decoherence.

\section{Is it plasma lensing?}\label{sec:scenarios}
\begin{figure}[!htb]
    \centering
    \includegraphics[width=0.75\linewidth]{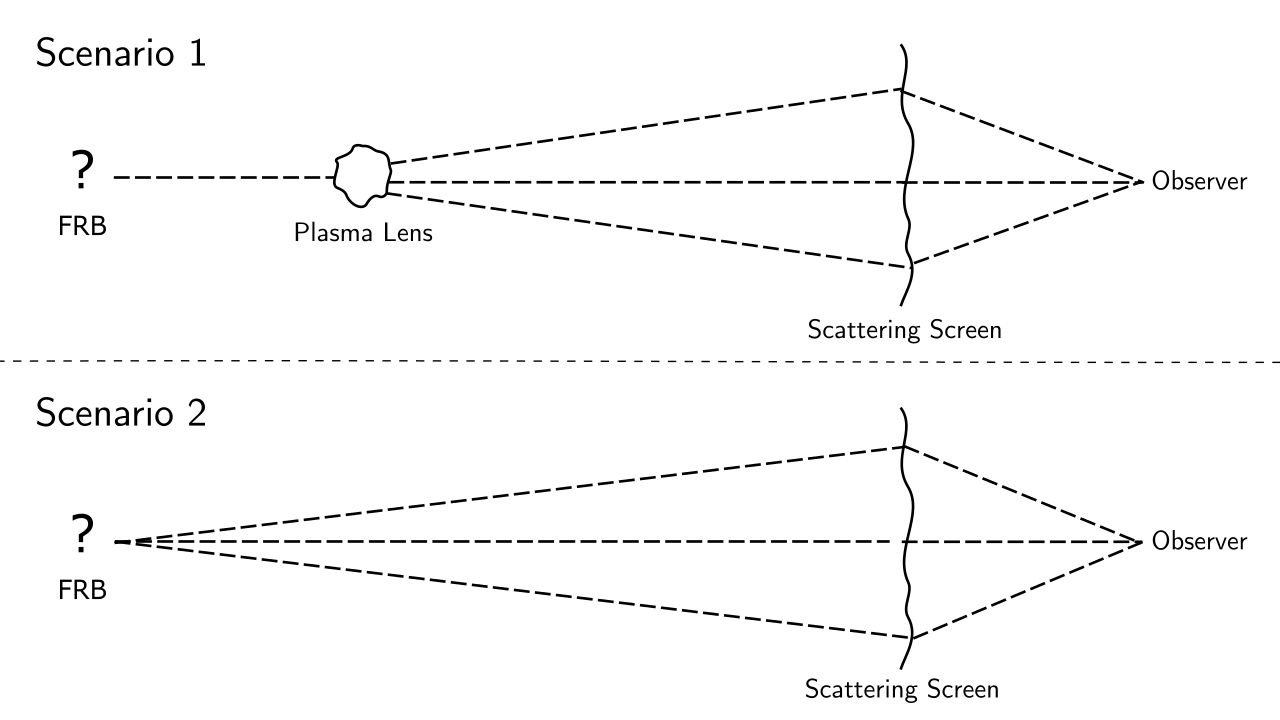}
    \caption{An illustration of the two possible scenarios that may explain the correlation signatures and morphology of FRB 20220413B. Scenario 1 is where an FRB encounters a plasma lens along the propagation, envisioned to be in the host galaxy. Scenario 2 has no plasma lensing to generate the morphology, but rather it is something related to the intrinsic FRB emission process. Both scenarios have the FRB sufficiently far from the scattering screen to remain unresolved by it. In addition to this, scenario 1 has the plasma lens itself unresolved by the scattering screen. }
    \label{fig:lens_scenarios}
\end{figure}

In the previous sections, we found evidence to suggest the morphology of FRB 20220413B can be explained by a plasma lens model. We found that the observed phase variance originates from the propagation of the electric field through a common, unresolved scattering medium. This is not an unexpected outcome as FRBs are thought to originate from extragalactic distances, and we observed many FRBs to have scintillation present in their spectra \cite{Schoen2021,Sammons2023,Masui2015,Nimmo2025,Wu2024}. The time-separated components (C and D) have a signature in the time-lag domain, which suggests the possibility that plasma lensing may have created this morphology. However, with no indication of coherence in the phase, there is not sufficient evidence to suggest these components are image copies of the other components. There is only evidence to suggest that they have propagated through the same scattering screen. 

This presents two scenarios that can explain the observed properties of this FRB. We illustrate the different scenarios in figure \ref{fig:lens_scenarios}. The first scenario is where this morphology is produced from a singular burst emission that encounters a plasma lens, such as in the host galaxy that produces multiple images, and then a scattering screen in the Milky Way galaxy to produce the observed scintillation. This is the scenario where all the components are images of one initial burst emission with common amplitude and phase. The second scenario is the case where all the distinct morphological frequency-time structures observed, including the time-separated components, are generated from the intrinsic emission process. This scenario has the same scattering screen as the first scenario.

\subsection{Scenario 1: Plasma Lensing}
In the first scenario, we want a plasma lens model that is able to create the time-separated components and the bifurcation of the burst into components A and B. We fit these parameters in section \ref{sec:morphfit}. Using the fit results, table \ref{tab:modelfit}, and the scintillation bandwidth measurements, we construct a two-screen scenario to model the lensing and simulate the observed voltage profile using a coherent propagation simulation\cite{Kader2024}. Without any angular information, there exists a large degenerate parameter space on the location of the screens and their separations. We can only determine the geometric lens parameter, $\mu_g= (1+z_l)D_{ol} D_{os}\theta_c^2 c^{-1} D_{ls}^{-1}$, which is degenerate with the angular scale and distance.

There exists one constraint that we may use to envision a possible system: we know the propagation through the scattering screen is unresolved. We envision the scattering to occur within the Milky Way, so we place the screen at 1 kpc. Given the DM of this burst, we place it at a redshift of $z=0.1$ using the Macquart relation \cite{Macquart2020}. For phase coherence to hold, the total angular extent of the source must be smaller than the coherence scale of the lens \cite{Kader2024}. We estimate the coherence scale using the Fresnel angle, $\theta_F = \left(c D_{ls}\right)^{1/2}\left(f D_{os} D_{ls}\right)^{-1/2}$, for a scattering screen at 1 kpc and at 400 MHz. For simplicity, we place the plasma lens screen at 1 kpc from the FRB in the host galaxy. For the scattering screen, we obtain a Fresnel angle of $\sim32$ $\mu$arcseconds at 400 MHz. For coherence to hold, we impose that the plasma lens is smaller than this scale.

\begin{figure}[!htb]
    \centering
    \includegraphics[width=0.7\linewidth]{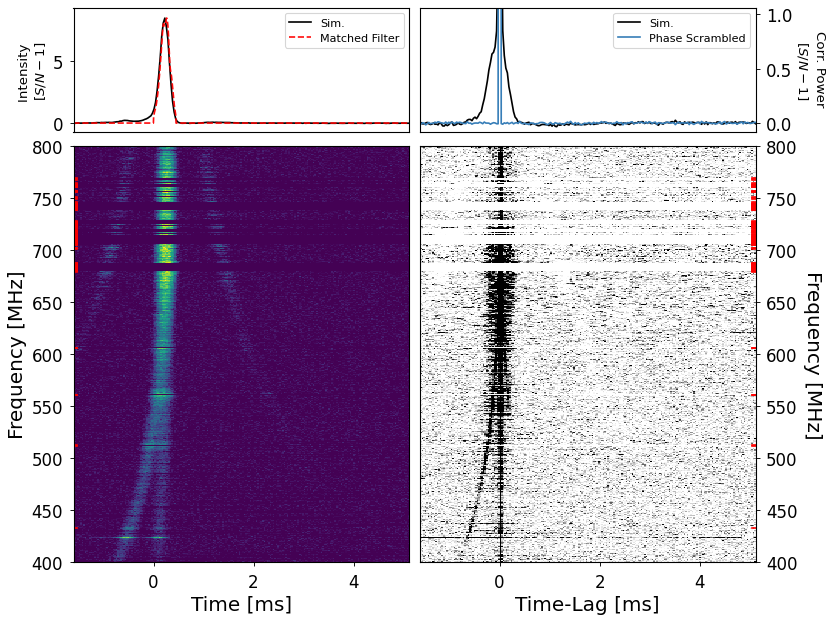}
    \caption{ The simulated baseband data and time-lag correlation for the fitted lens model scattered through a coherent scattering screen are shown. The right panels show the time-lag correlation, while the intensity and morphology of the simulation are shown in the left panels. The frequency average power is shown in the top panels. The simulated burst replicates the branching morphology of the burst and the correlation signature in the time-lag domain. Frequencies with caustics are masked and marked with red near the axis labels.}
    \label{fig:scen1_ftimecor}
\end{figure}

Using these parameters, we use the coherent simulation toolset from \textcite{Kader2024} to forward model the phase coherence of this lensing configuration. Figure \ref{fig:scen1_ftimecor} is the resulting simulation of baseband data. The branching morphology is replicated here in comparison to the real burst (figure \ref{fig:burst_ftimecorr}); however, the time-lag correlation provides a different signature. In this scenario, the plasma lens is coherent in nature, which produces the dispersive time-lag response in the right panel. We perform a search over DM in the time-lag domain to obtain the middle panel of figure \ref{fig:allscen_dmsearch}, similar to the search over DM for the real burst (figure \ref{fig:burst_dmsearch}). In contrast to the real burst, we find a localized response in this space at a $\Delta$ DM $\sim -0.05$ pc $\mathrm{cm}^{-3}$ and at a time-lag of $\sim -0.7$ ms. This is expected for a coherent phase delay. The lack of this response in the real burst implies that, if lensing did occur for this burst, then this phase response itself must also have decohered. There are many ways to decohere this response, but fundamentally, they all relate to the coherence scale of the plasma lens. The lack of large magnification present in the real burst, especially near the critical frequency, may itself be an indicator that decoherence has occurred \cite{Jow2022,Grillo2018}. For example, the angular extent of the FRB source may be resolved by a lens. As another example, we assume a simple, smooth model, but small-scale variations in DM near or on the lens plane may act like a secondary scattering screen that could decohere the phase response. Without any clear phase signature, decoherence is observationally the same as a phase incoherent burst emission process\cite{Main2021}. There is not enough information to allow us to distinguish between these two scenarios. We explore scenario 2 further in the next section.

\subsection{Scenario 2: Incoherent Emission}
To date, the FRB emission process is not known, and the large variations in morphology observed in FRBs \cite{Pleunis2021,Sand2025} indicated that the observed morphology for this event may simply be due to the emission process itself and without needing to invoke plasma lensing. Observation of phase coherence is the key to distinguishing between lensing and intrinsic emission processes; however, for this burst, we find no evidence for phase coherence. In this scenario, we will model the expected phase response for an incoherent emission.

We model the burst morphology using the DM for each component \cite{Faber2024} and impose the narrowband burst structure as a boxcar function applied in the frequency domain. For consistency, we use the burst parameters from the morphology analysis in scenario one. Explicitly, this means we assume no spectral dependence to the flux and use the fitted burst amplitude, $A$, and burst width, $\sigma$, to model the FRB (table \ref{tab:modelfit}). We model only four components of the burst: A, B, C, and D. 

\begin{figure}[!htb]
    \centering
    \includegraphics[width=0.7\linewidth]{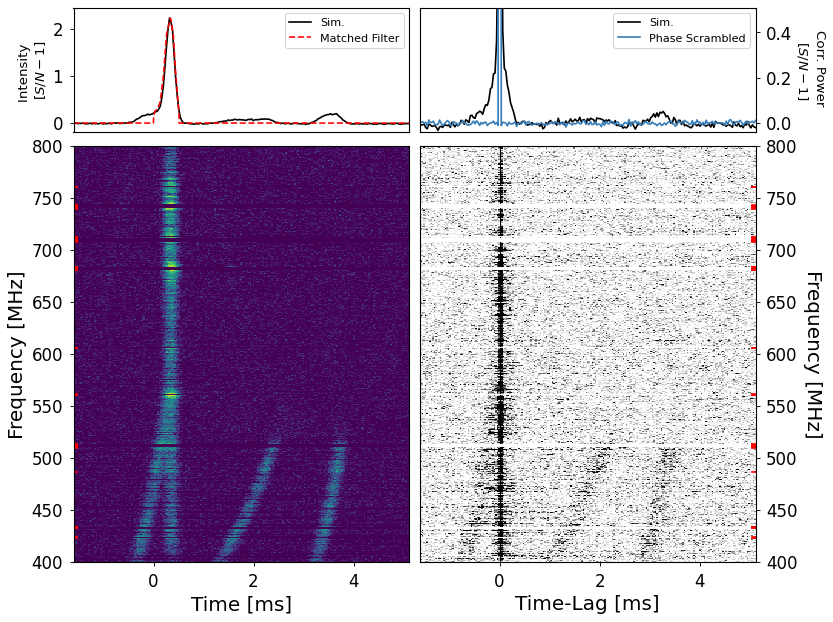}
    \caption{ The simulated baseband data and time-lag correlation for a phase incoherent FRB propagating through a coherent scattering screen are shown. The right panels show the time-lag correlation, while the intensity and morphology of the simulation are shown in the left panels. The frequency average power is shown in the top panels. The simulated burst replicates the time-lag correlation signature present in the real burst, indicating a common transfer function. Frequencies with caustics are masked and marked with red.}
    \label{fig:scen2_ftimecorr}
\end{figure}

We model the scattering response by applying a common propagation transfer function to all components. This creates scenario two, where there exists only a single common transfer function that produces the observed scintillation. Our scattering propagation function is the same as scenario one, such that the only change between the two scenarios is the existence of a plasma lens. That is, the scattering screen is placed 1 kpc from the observer while the source is located at a redshift of $z=0.1$. 

\begin{figure}[!htb]
    \centering
    \includegraphics[width=0.8\linewidth]{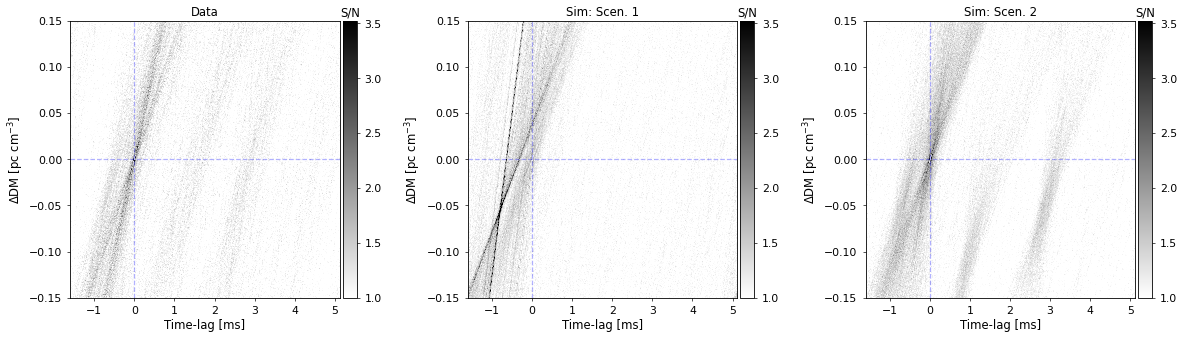}
    \caption{The search over DM in the time-lag domain for the data (left panel) and two simulated bursts (scenario 1 in the middle panel and scenario 2 in the right panel). The search is done over $400-500$ MHz with zero time-lag referenced to $400$ MHz. A coherent time-lag phase delay is a localized response in this parameter space. Scenario 1 is fully coherent, such that a localized response can be found. Scenario 2 has incoherent emission but a coherent scattering screen, resulting in an increase in S/N but no localized response. Comparing both scenarios to the data, scenario 2 is a better representation, indicating that there exists a common scattering screen, but the burst components are not intrinsically phase coherent with each other. }
    \label{fig:allscen_dmsearch}
\end{figure}

Figure \ref{fig:scen2_ftimecorr}  shows the intensity morphology of this burst in the left panels and the time-lag correlation of the burst in the right panels. The search over DM in the time-lag domain is shown in the right panel of figure \ref{fig:allscen_dmsearch}, where we find no localized response, unlike scenario one, shown in the middle panel of figure \ref{fig:allscen_dmsearch}. This is expected as the intrinsic response is incoherent, such that there should not exist a coherent phase response that would be localized to a specific DM and time-lag. We note the response here is similar to the response of the real burst, shown in the left panel of figure \ref{fig:burst_dmsearch}, indicating that there is no coherent phase delay in the burst. Furthermore, we note that a common scattering function can produce the observed increase in S/N in the time-lag correlation between burst components even when the components themselves are not intrinsically the same electric field. Both scenarios find a similarity in the correlation response, such that we may conclude there does exist a common scattering screen for all components of this FRB, but with no observed phase coherence, we cannot conclude whether or not lensing has occurred, only that there exists a coherent field propagation along some of the propagation path, but not along the full path.

\section{Conclusion}
FRB 20220413B shows a striking morphology that may be produced by propagation of an electric field through a plasma lens. We investigated this possibility in this work using the complex voltage data from CHIME. We performed a morphological fit to the intensity profile of the burst and found consistency with a plasma lensing scenario. We searched for phase coherence of the electric field for the burst by time-lag correlating the voltage data and found an excess in S/N for time-separated burst components. We analyzed the frequency spectra of four burst components through a frequency-lag correlation of the intensity spectra. We found the same scintillation is present in the spectra of the four burst components and is consistent with the expectations of scattering from the Milky Way. Two scenarios can explain the observed signatures of FRB 20220413B presented in this paper. The complex structure of this burst may be due to the intrinsic emission process, which propagated through a common scattering screen. If plasma lensing generated the complex morphology, then the coherence of lensing must have decohered before propagating through the common scattering screen.

The branching feature seen in the intensity profile of FRB 20220413B could be produced by plasma lensing near a caustic. We fit the morphology using a 1D Gaussian lens using catastrophe theory to constrain the parameter space and model the frequency-time intensity profile of the burst. We found consistency in the frequency-time structure of the burst, but not the flux. Our fit converged to a $\mathrm{DM}_{lens}\approx 0.1$ $\mathrm{pc}$ $\mathrm{cm}^{-3}$. The lack of an increase in flux due to the magnification near a caustic might indicate phase decoherence of the plasma lens or a more complex plasma lens.

We performed a time-lag correlation using a matched filter on the complex voltage data of the brightest component. We search over DM in the time-lag domain to search for a coherent phase response. We found evidence for an increase in S/N but no localized response in the DM search over the time-lag domain. This increase corresponds directly to the location of time-separated components. The lack of a localized, delta-like response indicates the burst components are not coherent images of each other.

We performed a scintillation analysis on the intensity spectra for four bright components of this burst. We correlated each component with itself and the other components. For the broadband component, a scintillation bandwidth of $\gamma_{scint}=71\pm 7$ kHz at 600 MHz with a scaling index $\alpha= 4.2\pm0.6$ is found. This is consistent with expectations of scattering and indicates scintillation is present for this burst. For the remaining components, we find an average scintillation bandwidth of $23\pm 2$ kHz at $450$ MHz, consistent with expectations from scattering through the Milky Way. This common scintillation bandwidth for all burst components indicates that the correlation response observed in the time-lag domain may result entirely from this common response; a plasma lens is not required.

We further investigate the possible scenarios that may produce this correlation signature by considering if there can exist a plasma lens that could generate this morphology, or if this may be generated from a phase incoherent FRB emission process. We use the fitted model parameters for the plasma lens and the measured scintillation bandwidth to create a fiducial lensing scenario that is fully phase coherent. We coherently propagate the phase through this two-screen lensing scenario to create the expected response for coherent plasma lensing. We additionally simulated a scenario where there is no lensing and therefore no phase coherence intrinsically, but there is still coherent propagation through a common scattering screen located in our galaxy. We find the time-lag response is consistent with the expectations for an incoherent emission rather than finding evidence for a coherent phase delay. We find that, while we are able to replicate the morphology using a plasma lens, without any phase coherence detected, we are unable to assert whether lensing has occurred or if this morphology is due to the FRB emission process. Nonetheless, this work shows the possibility that the complex burst morphologies of FRBs could be produced from propagation through inhomogeneous plasma acting as a lens. Note, these complex morphologies are not unique to FRBs, as pulsars, such as the Crab pulsar\cite{SerafinNadeau2025}, can show burst morphologies similar to the branching morphology observed here. The Crab emission is likely to be resolved by scattering screens \cite{Main2021} such that we expect an incoherent response, but other pulsars may present a similar phase signature \cite{Main2017} to that presented in this work. 

While we were unable to determine if plasma lensing occurred for this FRB, this work has presented a detailed study into determining methods to evaluate the scenario. We have presented key signatures that would indicate phase coherence and presented a method to fit for plasma lensed morphologies and extract lensing parameters in the frequency-time domain without angular information. We found evidence for partially coherent propagation having occurred for FRB 20220413B due to propagation through a common scattering screen. This work provides more evidence that coherent lensing is possible along the sight lines to FRBs. Decoherence from scattering is a large concern for any coherent lensing search. Scattering, however, is a frequency-dependent effect, and this work presents evidence that scattering through the Milky Way is able to maintain phase coherence between $400-800$ MHz. This suggests that while some lines of sight may not be coherent, others may be such that a coherent plasma or gravitational lens may be detected. For example, a sight line that is incoherent at the CHIME frequency band may be coherent at higher frequencies. This motivates a search for correlations in the voltage data of telescopes observing FRBs at higher frequencies. We note that a coherent lensing detection will directly be able to distinguish propagation effects from the intrinsic FRB emission processes. These lensing events present another method to determine key properties of the FRB emission, such as the size of the emission and the properties of plasma anywhere along the extragalactic propagation path.

\begin{acknowledgments}
\allacks
\end{acknowledgments}

\appendix
\section{Fitting a cusp caustic morphology}\label{sec:cuspfit}
The branching morphology of FRB 20220413B, under the assumption of lensing, can be presented as the caustic point where images merge. This is a critical point that can be used to constrain the parameters of a lens. In this section, we present an overview of how this morphology may arise using a simple plasma lensing to represent the morphology.
The Fermat potential is given by,
\begin{equation*}
    T(x_i;y,\mu_g,\mu_l)  = \mu_g\left( \frac{1}{2} (x_i-y_i)^2 + \alpha \Phi(x_i)\right),
\end{equation*}
where, through the semi-classical approximation of the KDI, $\mathbf{E} = \int d \mathbf{x} \mathbf{E}_s e^{i2\pi f T} $, we obtain the lensed images of the initial electric field \cite{Feldbrugge2019,Grillo2018}.
The gradient is given by,
\begin{equation}\label{eq:ferm_grad}
    \partial_x T(x;y,\mu_g,\mu_l)  = \mu_g\left(  x-y + \alpha \partial_x \Phi(x)\right),
\end{equation}
while the Hessian is given by,
\begin{equation}
    \partial_{xx} T(x;y,\mu_g,\mu_l)  = \mu_g\left(  1 +\alpha \partial_{xx} \Phi(x)\right).
\end{equation}

As a background to catastrophe theory, specifically the cusp catastrophe, let us consider a generic surface represented by the cusp potential \cite{Saunders1980},
\begin{equation}\label{eq:cusp_pot}
        \phi_{cusp}(w;\nu_1,\nu_2) = \frac{1}{4} w^4 + \frac{1}{2} \nu_2 w^2 +\nu_1 w,
\end{equation}
where $w$ parameterizes surface and there are two control parameters, $\nu_1$ and $\nu_2$. The gradient of this potential is a cubic function,
\begin{equation}\label{eq:cusp_grad}
       \partial_w \phi_{cusp} =  w^3 + \nu_2 w +\nu_1,
\end{equation}
and the Hessian is given by the polynomial,
\begin{equation}\label{eq:cusp_hess}
       \partial_{ww}\phi_{cusp} =  3w^2 + \nu_2 .
\end{equation}
When both $  \partial_w \phi_{cusp}=0$ and $ \partial_{ww}\phi_{cusp}=0$, the critical set is obtained for parameters $\nu_1$ and $\nu_2$, given by,
\begin{equation}\label{eq:cusp_crit}
           0 =   27 \nu_1^2  + 4  \nu_2^3 .
\end{equation}
When the control parameters satisfy equation \ref{eq:cusp_crit}, we have a caustic in our cusp potential. This description is a generic and topological one, so we are required to morph our lens model into this framework. Let's consider our 1D plasma lens that is Gaussian function \cite{Clegg1998},
\begin{equation}
   \Phi(x) = \exp(-\frac{x^2}{2}),
\end{equation}
with a gradient given by,
\begin{equation}\label{eq:gauss_grad}
    \partial_x \Phi(x) = -x\exp(-\frac{x^2}{2}).
\end{equation}
A plasma lens being represented by a cusp potential may be a generic feature of this lensing phenomenon \cite{Jow2023,Feldbrugge2019}. For this lens model, 1 to 3 images can form such that the cusp potential is a natural choice\cite{Jow2023}. Given our lensing potential, we require the gradient of the potential to be described by a cubic polynomial in order to obtain the control parameters. However, to ensure higher order accuracy in the roots of the gradient, we do not use a Taylor expansion about $x=0$ but rather the Pade approximant of order $(m,n)=(2,2)$ for a Gaussian function\cite{Baker1996} is,
\begin{equation}
       \exp\left(-\frac{x^2}{2}\right)  = \frac{1-\frac{x^2}{4}}{1+\frac{x^2}{4}} +O(x^6) .
\end{equation}
Using this Pade approximant with equations \ref{eq:ferm_grad} and \ref{eq:gauss_grad} we obtain the cubic equation,
\begin{equation}
\partial_x \Phi(x) = 0 = x^3 -\frac{y}{1+\alpha} x^2 +\frac{4\left(1- \alpha \right) }{1+\alpha}x -\frac{4y}{1+\alpha}.
\end{equation}
We then convert the cubic into a depressed cubic to obtain the cusp control parameters, where
\begin{equation}\label{eq:cuspparams}
\begin{split}
    \nu_2 & =  \frac{12- 12\alpha^2  -y^2}{3(1+\alpha)^2} \\
        \nu_1 & = -y \frac{72 + 144 \alpha^2 +216\alpha +2y^2  }{27 (1+\alpha)^3} .\\
\end{split}
\end{equation}
The caustics for this lens model are then found using the control parameters, given by equation \ref{eq:cuspparams}, and asserting equation \ref{eq:cusp_crit} is satisfied. As an example, there is a cusp catastrophe when $\nu_1 = \nu_2=0$. In this case, the physical lens parameters for this lens model satisfy, 
\begin{equation}\label{eq:cuspparamscrit}
\begin{split}
        \nu_2 & = 0  =  12- 12\alpha^2  -y^2 \\
        \nu_1 & =  0 = -y \left(72 + 144 \alpha^2 +216\alpha +2y^2  \right) .\\
\end{split}
\end{equation}
If $y= 0 $ the source, lens, and observer are coaligned, and then if the lens strength parameter, $\alpha=\pm1$, we have a cusp catastrophe in the Fermat potential.

\begin{figure}[!htb]
    \centering
    \includegraphics[width=0.7\linewidth]{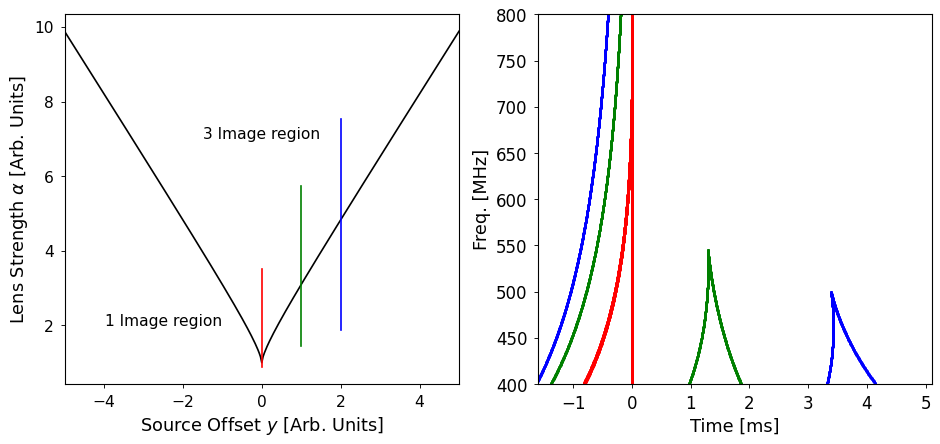}
    \caption{The parameter space mapping the 1D Gaussian plasma lens to a cusp potential is shown in the left panel. The right panel shows the frequency and time range over which FRB 20220413B is observed in the CHIME band. The black line in the left panel shows the critical curve dividing the regions with different numbers of images. The mapping of images from parameter space to the frequency-time domain is shown with the colored lines. The red line is the best fit for FRB 20220413B. The green and blue lines change in offset and critical frequency to produce a morphology resembling the time-separated components of the burst.  }
    \label{fig:ftime_imgmorph}
\end{figure}

We now have a mapping of our physical lens parameters to the topological control parameters. Importantly, we are able to determine the parameters that will generate a caustic. We may now constrain and fit these parameters by asserting that there exists a caustic in the morphology of FRB 20220413B. For our lensing model, $\alpha = k_{\mathrm{DM}} \mu_g^{-1} \mathrm{DM}_{lens} f^{-2}$. We assert the caustic point lies within $400-800$ MHz. We have a source offset parameter $y$ which provides a critical $\alpha$, found by using equations \ref{eq:cusp_crit} and \ref{eq:cuspparams}. Given the critical frequency $f_c$, and the critical lens strength parameter, $\alpha(y,f_c)$, we can obtain $\mu_g$ by allowing $\mathrm{DM}_{lens}$ to vary. The Fermat potential is then fully described by parameters $f_c$, $y$, and $\mathrm{DM}_{lens}$. By allowing these parameters to vary, we fit the frequency-time intensity profile of the FRB by asserting that there exists a caustic point. Figure \ref{fig:ftime_imgmorph} presents the control parameter space for the cusp potential in the left panel. The black line indicates the caustic curve where parameters that lie on this curve will generate a caustic. The colored lines represent the mapping from this parameter space to the frequency-time profile for the images, shown in the right panel. All images are referenced to the DM of the main image at $y=0$, represented in red. We note that this branching morphology is similar to that of FRB 20220413B, while the morphologies generated through a non-zero source offset create structures similar to those seen at later times for FRB 20220413B. 

We caution that we fit for the main branching component in this work, but fitting for all components requires additional lensing components to the model. For example, to account for them, we require additional Gaussian components in the lensing potential. These add additional parameters such as offset positions, size, and strength, which will alter the lensing model and, importantly, may require higher-order catastrophe potentials to describe. We, therefore, choose to fit only for the main branching component as we seek to assert whether plasma lensing can explain the morphologies of FRB 20220413B and not what lens model is the best fit for this event. Nonetheless, the morphologies presented in figure \ref{fig:ftime_imgmorph} show that all the components could be modeled by propagation of the electric field through a plasma lens.

%\printbibliography
\bibliographystyle{apsrev4-2}
\bibliography{biblio.bib}% Produces the bibliography via BibTeX.

\end{document}